# Growth dynamics and gas transport mechanism of nanobubbles in graphene liquid cells

Dongha Shin[1]‡, Jong Bo Park[1]‡, Yong-Jin Kim,[1,2] Sang Jin Kim[1], Jin Hyoun Kang[1], Bora Lee, Sung-Pyo Cho[1], Byung Hee Hong[1] and Konstantin S. Novoselov[2]

*[1]Department of Chemistry, College of Natural Science, Seoul National University, Seoul 440-746, Korea. [2]Department of Physics & Astronomy, University of Manchester, Manchester M13 9PL, UK*

‡These authors contributed equally to this work.

**Formation, evolution, and vanishing of bubbles are common phenomena in our nature, which can be easily observed in boiling or falling waters, carbonated drinks, gas-forming electrochemical reactions, etc[1,2]. However, the morphology and the growth dynamics of the bubbles at nanoscale[3-10] have not been fully investigated owing to the lack of proper imaging tools that can visualize nanoscale objects in liquid phase. Here we demonstrate, for the first time, that the nanobubbles in water encapsulated by graphene membrane can be visualized by *in situ* ultrahigh vacuum transmission electron microscopy (UHV-TEM)[11-13], showing the critical radius of nanobubbles determining its unusual long-term stability as well as two distinct growth mechanisms of merging nanobubbles (Ostwald ripening and coalescing) depending on their relative sizes. Interestingly, the gas transport through ultrathin water membranes at nanobubble interface is free from dissolution, which is clearly different from conventional gas transport that includes condensation, transmission and evaporation[14]. Our finding is expected to provide a deeper insight to understand unusual chemical, biological and environmental phenomena where nanoscale gas-state is involved.**

There have been intensive efforts to characterize the nanobubbles in liquid phase, which includes ion conductance measurement through a solid-state nanopore[15], topographic imaging by atomic force microscopy (AFM)[16] and direct visualization by optical methods[17,18]. None of these, however, was capable of imaging the liquid phase nanobubbles in real time with sub-10 nm resolution. In this regard, *in-situ* TEM would be the best method to observe the behaviours of nanobubbles, but the resolution is still limited by the thickness and the robustness of liquid cell membranes. Recently, it was reported that graphene can be utilized as a perfect liquid cell membrane for *in-situ* TEM imaging of nanocrystal growth thanks to its atomic thickness, flexibility, extraordinary mechanical strength and high conductivity[11]. Thus, we tried to investigate the evolution of nanobubbles by encapsulating them in a graphene liquid cell membrane for *in-situ* TEM imaging in ultra-high vacuum.

The graphene liquid cell was fabricated by the sequential wet transfer of monolayer graphene synthesized by chemical vapour deposition (Fig. S1)[19-21]. The water islands are naturally captured during the wet transfer process of graphene to a graphene-supported TEM grid (Fig. S2). As shown in Fig. 1, the top and side views of nanobubbles show the plano-convex morphology whose diameter ranges from 5 to 15 nm. It should be noted that the high mechanical flexibility and strength of graphene allows the cross-sectional imaging of nanobubbles in a folded liquid cell (Fig. 1e to g). We found that the contact angles of nanobubbles are ~71.2±1.2° regardless of their sizes (Fig. S3). These values were used to calculate the internal pressure of nanobubbles using Young-Laplace equation, $\Delta P = 2\gamma / R_c$, where $\Delta P$ is the pressure difference across the nanobubble interface, $\gamma$ is the surface tension of water, and $R_c$ is the curvature radius of nanobubbles (Fig. 1h). According to this equation, Young-Laplace pressure inside a 10 nm-diameter nanobubble is calculated to be 27 MPa, which is 270 times higher than ambient pressure.

According to classical diffusion theory, the lifetime of a nanobubble was predicted to be ~1 μs[22]. In fact, however, nanobubbles are very stable even for several hours as revealed by liquid-phase AFM[23]. Many explanations on this superstability of nanobubbles were proposed, including stabilization by three-phase contact line pinning[24] and dynamic equilibrium at water-vapour interface[25,26]. In addition, the critical radius of stable nanobubbles was predicted to be ~1.7 nm by molecular dynamic (MD) simulation[27] and ~85 nm by dynamic equilibrium theory[26], but there has been no experimental confirmation so far. Here we show, for the first time, that the critical radius of stable nanobubbles is 4~6 nm as shown in Fig. 2a and b. For the nanobubble radius below the critical range, the radius keeps decreasing until it completely collapses, while the nanobubble lager than ~6 nm persists for more than 10 min. Surprisingly, the model calculation based on the structural parameters from the TEM observation gives the stable radius of 6.10 nm, which fits the critical radius range in Fig. 3a and b (please see Supplementary Information for more details).

Nanobubbles are growing by merging with adjacent nanobubbles, which shows clearly different two pathways depending on their relative sizes. In case that the sizes are distinctively different ($R>R'$), the smaller bubble tends to disappear near the surface of the growing larger bubble (Fig. 2c), which is similar to Ostwald ripening that is known as a solid state phenomenon that small crystals are dissolved and redeposited on to the surface of larger crystals. It seems that gas diffuses from one bubble to another across the persisting boundary. On the other hand, two similar-sized nanobubbles ($R\sim R'$) show a coalescing process after breaking their interface, followed by reshaping into dumbbell-like and spherical morphology (Fig. 2d).

Fig. 3a and b show that there exists critical radius range for the stability of nanobubbles. The nanobubbles whose radii larger than 6 nm persist more than 10 min, while smaller bubbles tend to disappear in 1 min. In the Ostwald ripening-like process, the radius of the smaller nanobubble show a change in slope with respect to time, while the radius of the larger bubble steadily increases (Fig. 3c). Here, we suppose that a new pathway of gas diffusion is created when the thickness of the interface is smaller than ~2.3 nm, where the instantaneous rupture of the interface allows the massive diffusion from a highly pressurized smaller bubble to a larger bubble. We define it as 'direct gas diffusion (or transport)' to be distinguished from 'indirect gas diffusion'. (See Supplementary Movie 3). The internal pressure of the small nanobubble increases from 140 MPa to 400 MPa as shown in Fig. 3d, which is driving force for gas transport from the small bubble to the large bubble.

Fig. 4 shows the two different pathways of gas transport (Fig. 4a) and time-resolved TEM section analysis of Ostwald ripening nanobubbles (Fig. 4b). Usually, conventional gas transport between remote nanobubbles includes condensation, transmission and evaporation steps[14]. However, in case that two Ostwald ripening nanobubbles come into contact each other, the gaseous particles seem to diffuse as a discrete packet from one to another through the ultrathin water membrane without hydration, which needs to be importantly considered for the assembly and function of biomolecules and other systems where nanoscale gas state is involved. The instantaneous breakjunction of the ultrathin water membrane appears dominantly as the thickness decrease below ~2 nm as shown in Fig. 4b.

In summary, the liquid phase nanobubbles encapsulated by graphene membrane were visualized by *in-situ* UHV-TEM, showing the critical radius of nanobubbles determining its long-term stability as well as two different growth processes of merging nanobubbles depending on their relative sizes. It is remarkable that the instantaneous rupture of the ultrathin water membrane between nanobubbles allows direct unhydrated gas transport that has not been observed so far. We believed that this phenomenon needs to be importantly considered in various biological and environmental systems where nanoscale gas state is involved.

## Methods

**Preparation of monolayer graphene.** Graphene was synthesized by the chemical vapour deposition method on a high purity copper foil (Alfa Aesar, 99.999%) with flowing 70 mTorr $H_2$ and 650 mTorr $CH_4$ gas. As grown graphene on Cu was spin-coated with PMMA (poly methyl methacrylate) and back-side graphene was etched using oxygen plasma. Then, the PMMA layer on graphene was removed by acetone. Remaining copper was etched in 1.8wt% ammonium persulfate (APS) solution. Finally, the monolayer graphene was rinsed with distilled water several times.

**HR-TEM observation of nanobubbles.** Electron microscopic analysis was carried out using *in-situ* UHV-TEM (JEOL, JEM 2010V) operated at 200 keV[12,13]. Its point resolution at Scherzer defocus is 0.23 nm and lattice resolution is 0.20 nm. The ultimate base pressure in the chamber was less than $2 \times 10^{-10}$ Torr, and the pressure during observation was below $5 \times 10^{-9}$ Torr. All the experiments shown here were performed at room temperature. The UHV *in situ* HRTEM observations were employed optimized parameters for imaging, i.e. there were recorded close to the Scherzer defocus and the sample height was adjusted to keep the objects focused in the optimum lens current, because HRTEM images often change depending on the high beam current density and defocus. *In-situ* real-time HRTEM images were recorded by a digital video recorder at the time resolution of 1/30 s equipped with an on-line TV camera system (Gatan model 622SC). The typical electron beam current density at the specimens was a very small value of ~1 A cm$^{-2}$. It is well known that an electron beam can adversely affect irradiation damages of a sample during examination in an EM (e.g., heating, electrostatic charging, ionization damage, displacement damage, sputtering and hydrocarbon contamination)[28]. However, the above-mentioned observation conditions, especially a very small current density and an UHV situation reduced the risk of irradiation damages and hydrocarbon contaminations to the minimum. Although atomic resolution of the JEM 2010V with a LaB6 filament used in this study as compared to that of an EM with a field-emission gun (FEG) filament falls, its current density is lower by about 100-1000 times than that of the FEG filament. Moreover, the current density of ~1 A cm$^{-2}$ at most brings a temperature increase of a few degrees of celsius[28], which perhaps hardly influences the sample in a recoding time, usually 2 to 5 minutes. In fact, while observing the magnified images, no changes in image detail arising from electron beam irradiations were detected. Therefore, we believed that these advantages as well as unique capabilities of graphene liquid cell as a perfect membrane for EM imaging[11,29] has enabled the characterization of nanobubbles without contamination in this study.

Received (date)

**Supplementary Information** accompanies the paper

**Acknowledgements** This research was supported by the Basic Science Research Program (2011-0006268, 2012M3A7B4049807), Converging Research Centre Program (2013K000162), the Global Frontier R&D Program on Centre for Advanced Soft Electronics(20110031629) and the Global Research Lab (GRL) Program(2011-0021972) through the National Research Foundation of Korea funded by the Ministry of Science, ICT & Future, Korea. This research was also supported by Inter-University Semiconductor Research Centre (ISRC) at Seoul National University. We also thank kind technical support from JEOL Inc. in HAADF STEM (JEM ARM 200F with a cold type field-emission gun) imaging of graphene.

**Authors Contribution** B.H.H., S.-P.C. and K.S.N. conceived and supervised the project. D. S. and J.B.P. led the project. B.H.H., S.-P.C., D.S., J.B.P., Y.-J. Kim wrote the manuscript. S.-P.C. carried out TEM imaging and analysis. J.B.P., S.J.K., Y.-J. Kim prepared the samples. J.B.P. and D. S. assisted with theoretical modelling. J.H.K. and B.L. assisted with surface analysis. Y.B.H. and J. H. P. helped with fluidic analysis.

**Authors Information** Reprints and permissions information is available atwww.nature.com/reprints. Correspondence and requests for materials should be addressed to B.H.H. (byunghee@snu.ac.kr) and J.-S.-P. (chosp@snu.ac.kr).

**FIGURE LEGENDS**

**Figure 1. Morphology of nanobubbles in water trapped between two single layered graphene sheets. a-d**, A graphene liquid cell fabricated on a flat TEM grid (copper or molybdenum) showing the top views of nanobubbles. **c** and **d**, *In-situ* snapshot images of nanobubbles obtained by ultra-high vacuum (UHV) TEM (200 keV, ~ 5 x $10^{-9}$ Torr). Scale bars, 10 nm. **e-g**, A folded graphene liquid cell showing the side views of nanobubbles. The contact angles are measured to be 71.2±1.2°. Scale bars for f and g, 10 nm and 5 nm, respectively. **h**, The schematic image of a nanobubble on solid surface and its structural parameters including surface radius ($R$), contact angle ($\theta_C$), curvature radius ($R_C$), and height ($H$). The full movie is available in Supplementary Information (Movie S1).

**Figure 2. Snap shots of TEM images showing the time evolution of different kinds of single and double nanobubbles. a** and **b**, The snap shots of TEM images showing the vanishing and stable nanobubbles, respectively. The nanobubbles smaller than critical radius tend to shrink with time and disappear in ~40 sec, while the larger bubbles persist for more than 10 min. Scale bars, 5 nm. The full movie is available in Supplementary Information (Movie S2). **c** and **d**, The snap shots of TEM images showing the merging of adjacent two nanobubbles observed for 15 and 50 seconds, respectively. When the nanobubble sizes are significantly different, it shows an Ostwald ripening like merging process, while the similar-sized bubbles are coalescing as their inter-bubble boundary breaks. Scale bars, 10 nm. The full movie is available in Supplementary Information (Movie S3).

**Figure 3. Analysis of TEM images of single nanobubble and Ostwald ripening nanobubbles. a** and **b**, Average radius and internal pressure changes of vanishing and stable nanobubbles with time, respectively. The pressure was calculated by Young-Laplace equation. This result indicates that the critical radius of nanobubbles lies between 4~6 nm. **c,** Time evolution of radius of growing (red), vanishing (blue) nanobubble and inter-bubble distance measured in **Fig. 2c**. **d**, Calculated internal pressure of Ostwald ripening nanobubbles in **Fig. 2c**. The inset shows the calculation result representing the liquid water density with respect to their relative size and distance between two adjacent nanobubbles, indicating that the water density decreases at the interface region as two bubbles get closer, which is a driving force to put two remote bubbles together.

**Figure 4**. **Direct and indirect gas transport in Ostwald ripening nanobubbles. a**, Schematic representations explaining indirect and direct gas transport. Between two remote nanobubbles, gases are slowly transported through conventional condensation, transmission and evaporation processes, while interfacial nanobubbles show direct gas transport through the ultrathin water membrane without hydration. **b**, Time-resolved TEM section analysis of the interbubble region (line a-b), extracted from Supplementary Movie 3. The thickness of water layer gradually decreases with time, and the occurrences of instantaneous breakjunctions are clearly observed as indicated by white arrows. Opening and closing of water membrane are clearly visible between 11 and 13 sec.

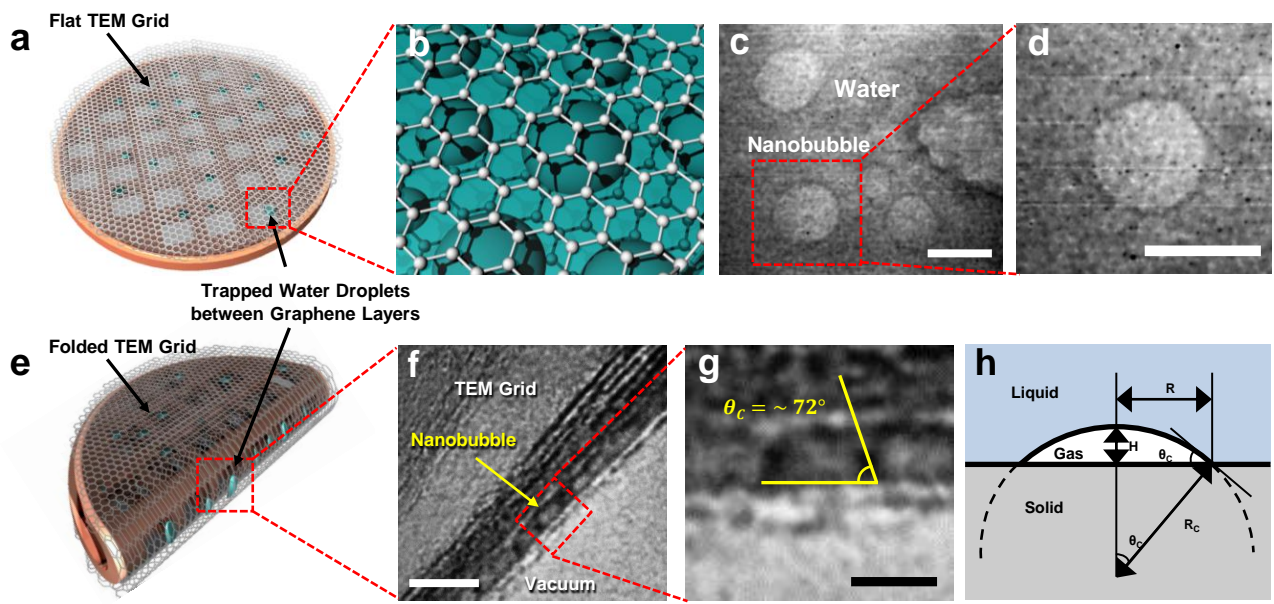

**Figure 1**

**a** **Vanishing Nanobubble**

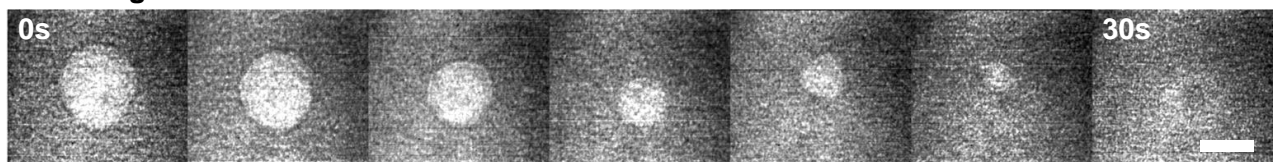

**b** **Stable Nanobubble**

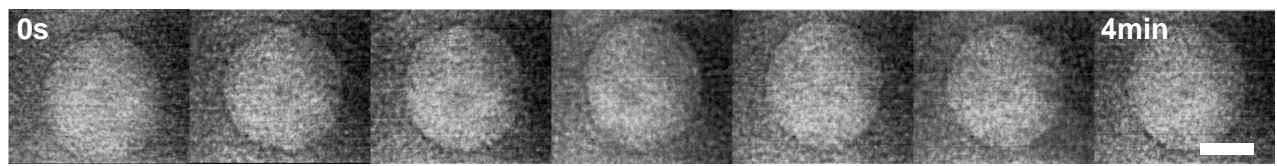

**c** **Ostwald-Ripening Nanobubbles**

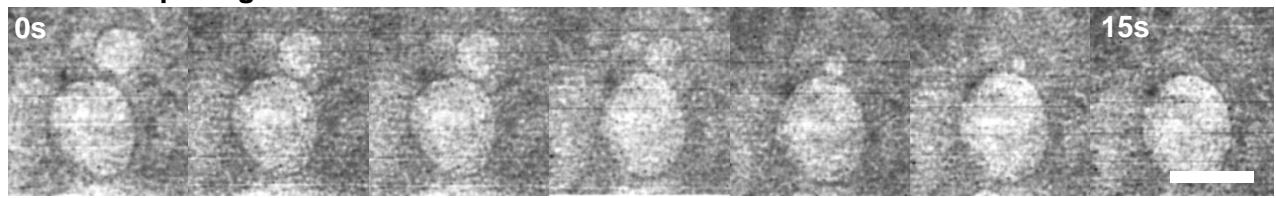

**d** **Coalescing Nanobubbles**

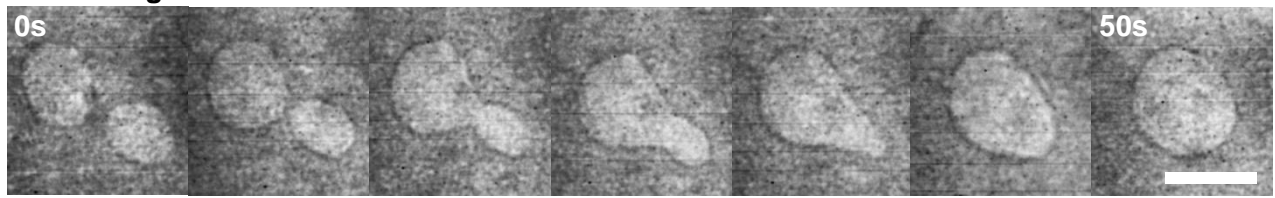

**Figure 2**

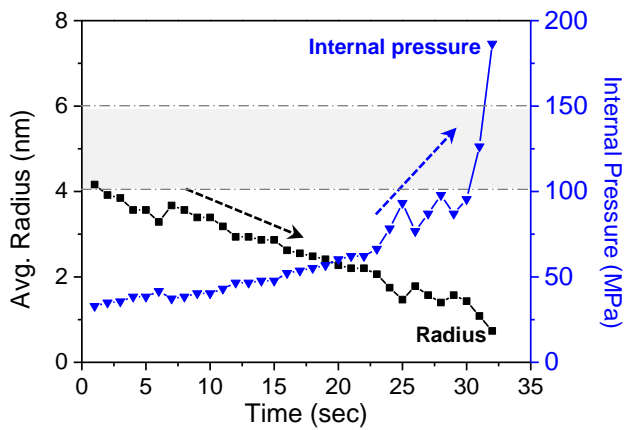
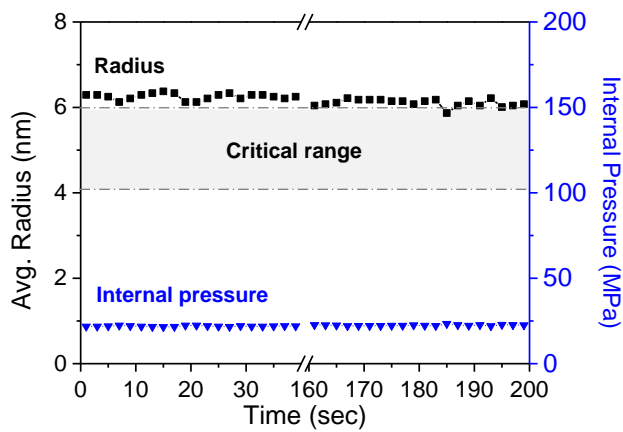
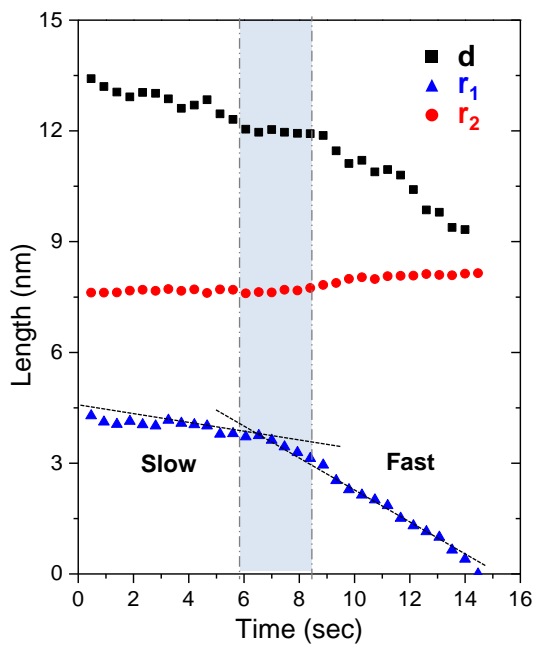
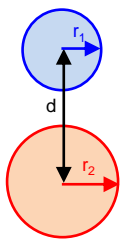
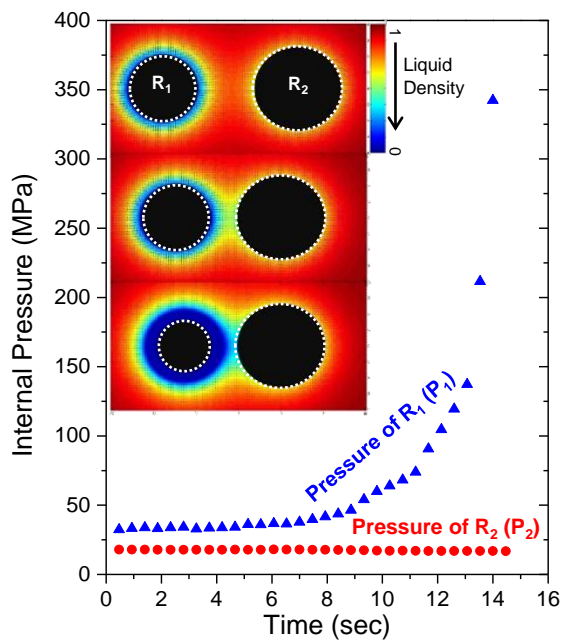

**Figure 3**

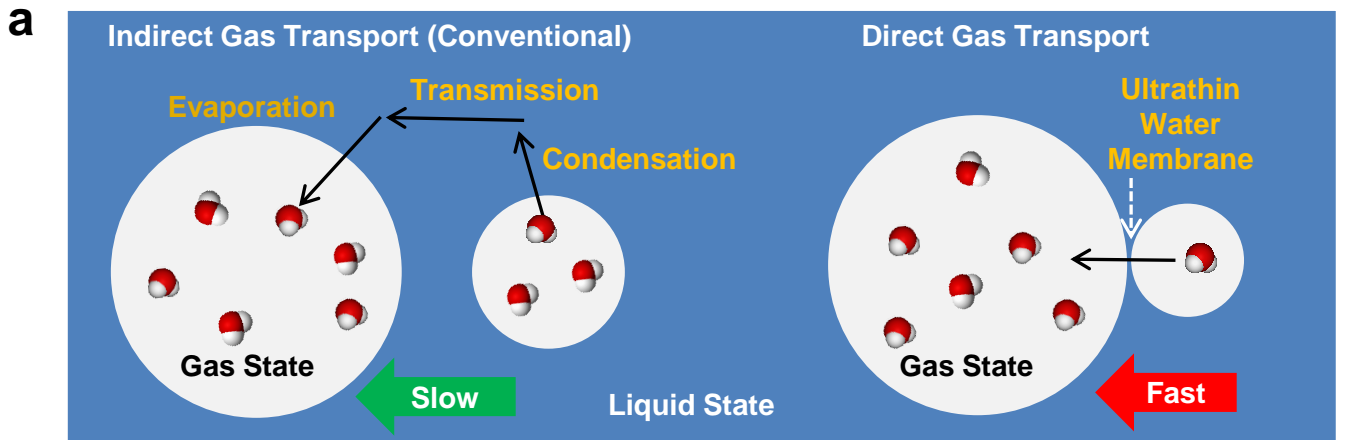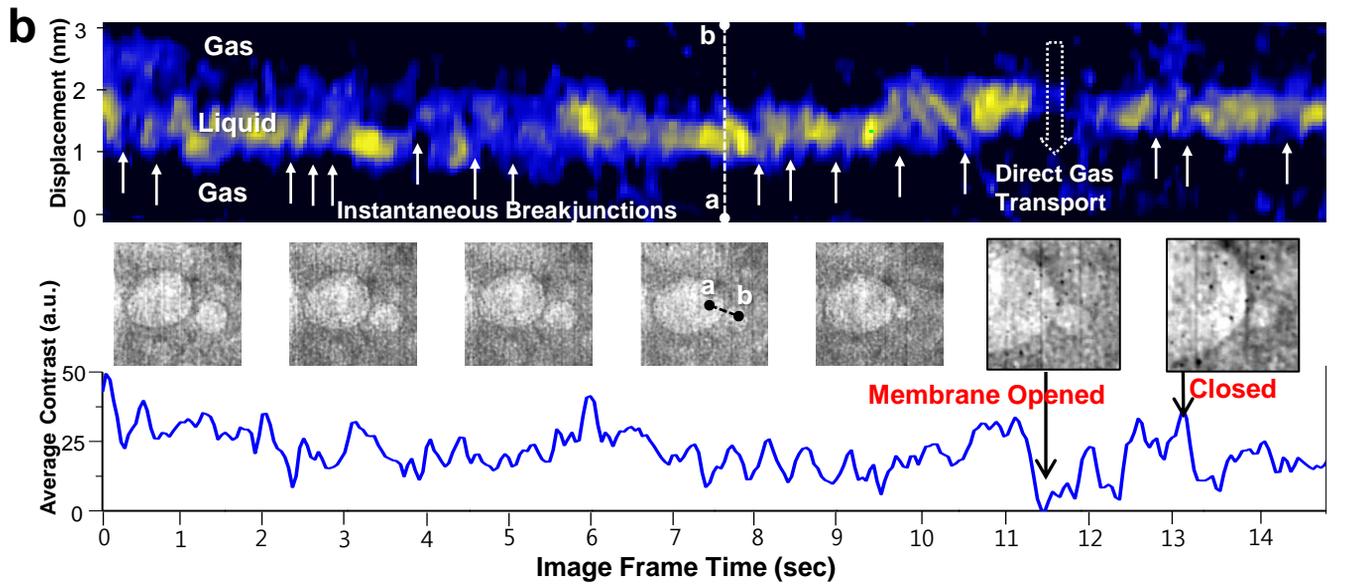

**Figure 4**

# Supplementary Information

## A. Calculation on the stable radius of plano-convex shaped nanobubbles

The stable radius of nanobubbles was calculated considering the structural parameters from TEM observation and the molecular dynamics simulation results by Matsumoto *et al.* (Ref. 27). The setting temperature of water is 300K, and the system volume was fixed at $V$ = 30 x 30 x 7.5 (nm)$^3$. The liquid pressure of system $P_{sys}$ can be estimated from the density of liquid $\rho_{liq}$ as

$$P_{sys} = A \frac{mN}{V - V_{bubble}} + B$$

where $m$ is the molecular mass of water, $N$ is the number of molecules, and $V_{bubbule}$ is the volume of nanobubble. A and B are the constants determined approximately by a linear function of $P_{sys}$ with respect to $\rho_{liq}$. Considering the plano-convex shape of nanobubbles, $V_{bubbule}$ was calculated by simple integral as

$$V_{bubble} = \int_{R_c - H}^{R_c} (R_c^2 - y^2)\pi dy = \pi R^3 \left\{ \frac{\left(\tan\frac{\theta_c}{2}\right)^2}{\sin\theta_c} - \frac{1}{3}\left(\tan\frac{\theta_c}{2}\right)^3 \right\}$$

The surrounding liquid pressure of nanobubble, $P_{liq}$ is given by $P_{liq} = P_{vap} - \Delta P$, where $P_{vap}$ is the gas pressure inside the nanobubble and $\Delta P$ is Young-Laplace pressure. At 300K, the vapor density inside nanobubble is very low, so it can be set as $P_{vap} = 0$ in our calculation. Thus, $P_{liq}$ simply can be expressed as $\sim -\Delta P$. Now the radius of stable nanobubble can be derived from the equilibrium equation between liquid and system pressure, $P_{liq} = P_{sys}$ as following:

$$-\frac{2\gamma \sin\theta_c}{R} = \frac{AmN}{6750 - \pi R^3 \left\{ \frac{\left(\tan\frac{\theta_c}{2}\right)^2}{\sin\theta_c} - \frac{1}{3}\left(\tan\frac{\theta_c}{2}\right)^3 \right\}} + B$$

The constant values are approximated as $A = 2 \times 10^{24}$, $B = -1980$, $\gamma = 71.97$ mN/m, $N = 211,003$ and $\theta c = 72°$. $A$ and $B$ are adopted from the MD simulation by Matsumoto *et al.* (Ref. 27), $\theta c$ is determined based on the TEM results, and $\gamma$ is the surface tension of water. In this way, the stable radius of a nanobubble was calculated to be 6.10 nm.

## B. Supplementary Figures

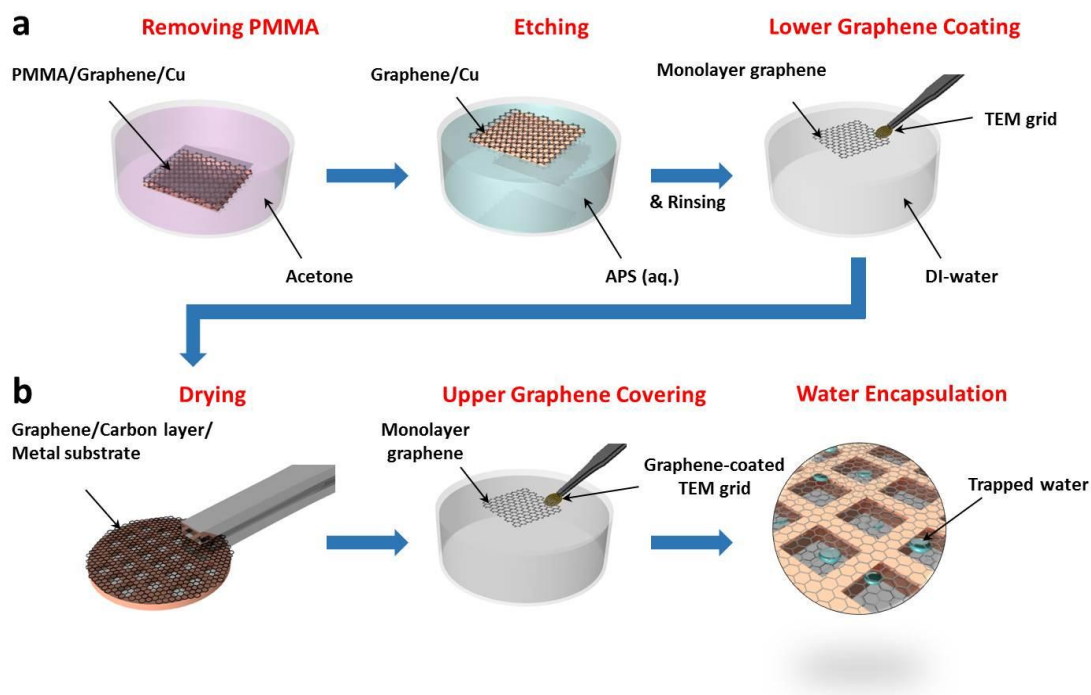

**Figure S1. Schematic representation for the preparation methods of graphene liquid cells. a**, Fabrication of graphene coated TEM grid. The monolayer graphene without PMMA support was prepared by removing PMMA with acetone, followed by Cu etching with 1.8wt% ammonium persulfate (APS) solution. Finally rinsing and transferring complete the graphene-supported TEM grid. **b**, Capturing water islands by transferring the second layer graphene. After drying, another monolayer graphene floating on water was transferred on to the graphene-supported TEM grid, where residual water on graphene can be trapped naturally between two graphene layers.

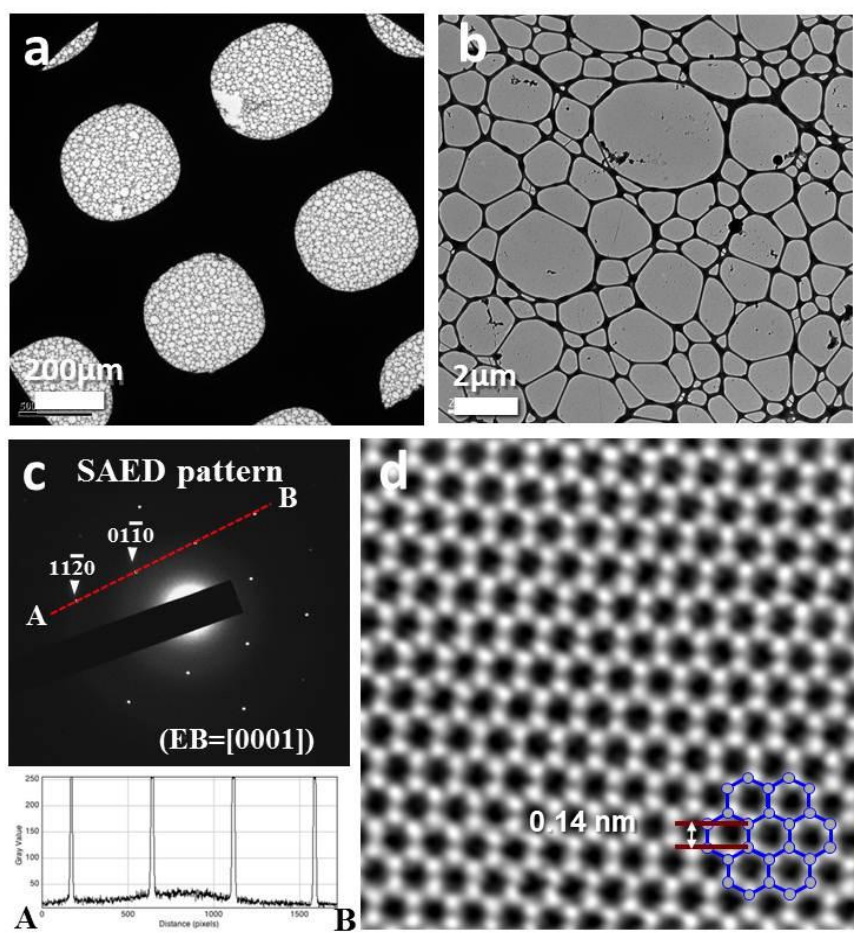

**Figure S2. Characterization of high quality CVD graphene using HR-TEM and HAADF-STEM. a** and **b**, TEM images of monolayer graphene on a Cu grid with amorphous carbon as support layer. **c**, Selected area electron diffraction (SAED) pattern of graphene monolayer, evidenced by the same intensity along the A-B line profile. **d**, Atomic resolution image of high quality graphene obtained by high-angle annular dark filed (HAADF) scanning transmission electron microscope (STEM).

4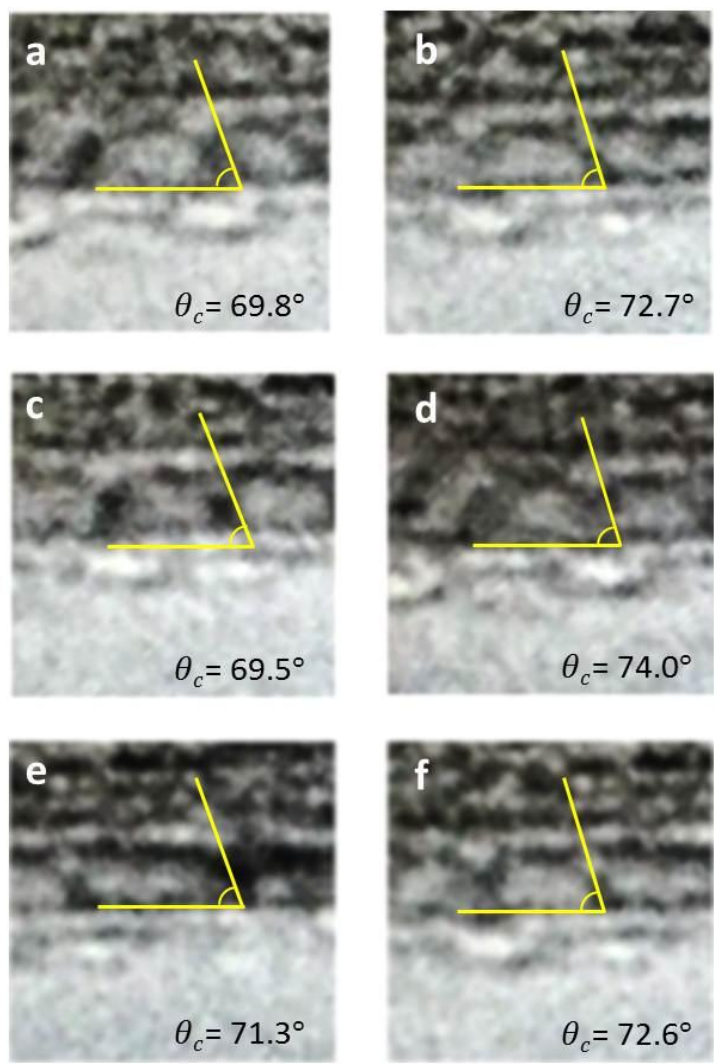

**Figure S3. Cross-sectional TEM images of nanobubbles in a graphene liquid cell. a-f**, TEM images showing the contact angles of nanobubbles in water trapped between sandwiched graphene, ranging from 70 to 74° (see supplementary Movie S1). The average contact angle of the nanobubbles is ~ 72°. Scale bar, 5 nm.

\* Supplementary movies available on request: byunghee@snu.ac.kr